\documentclass{jltp}

\usepackage{graphicx} 

\title{The Role of Electron-Hole Symmetry Breaking in the Kondo Problems
\thanks{Dedicated to Peter Wolfle on the occasion of his 60th birthday 
who is an old-time friend of us and whose work has been always stimulating
for us.}}

\author{O. \'Ujs\'aghy$^a$, K. Vlad\'ar$^b$, G. Zar\'and$^{a,c}$, and 
A. Zawadowski$^{a,b}$}

\address{$^a$ Department of Theoretical Physics and Research Group \\of
the Hungarian Academy of Sciences, \\Technical University of Budapest,
H-1521 Budapest, Hungary\\
$^b$ Solid State and Optical Research Institute of the Hungarian
\\Academy of Sciences, H-1525 Budapest, Hungary\\
$^c$ Lyman Physics Laboratory, Harvard University, Cambridge MA 02145}

\runninghead{O. \'Ujs\'aghy \textit{et al.}}{The Role of Electron-Hole 
Symmetry Breaking in the Kondo Problems}

\newcommand{\ep}{\varepsilon}
\def\bsigma{\mbox{\boldmath$\sigma$}}

\begin{document}

\maketitle

\begin{abstract}
In studying the different Kondo problems it is generally assumed that
breaking the electron-hole symmetry does not affect the perturbative
infrared divergencies.
It is shown here that, in contrast, breaking that symmetry 
may in some cases lead to observable modifications
while in other cases it does not.

\end{abstract}

\section{INTRODUCTION}

The heart of the Kondo phenomena is in the infrared divergences or the
orthogonality catastrophe. Those phenomena occur in the neighborhood
of the Fermi energy. Thus, breaking the electron-hole symmetry on the
large energy scale of the conduction band width does not enter directly. 
In the first iterations of the scaling at large energy scale that may, 
however, induce non-universal effects. 

In the first part of this paper, the spin Kondo problem with
asymmetric bandwidth cutoff is studied. The results are somewhat unexpected, as
the generated potential scattering in the intermediate energy region plays a
relevant role in the scaling equations, which is not the case for
symmetric bands. 

Two examples are presented where the breaking of the
electron-hole symmetry leads to instabilities of the scaling
trajectories. The first example is a two-level impurity model with
commutative couplings where there is an additional potential
scattering at the impurity site which generates non-com\-mutative terms in
the Hamiltonian. 
The other one is a two-level system with two-channel Kondo behavior,
where the spin degeneracy is broken by spin-orbit scatterings in the
neighborhood of the impurity atom. In both cases the energy splitting
or direct tunneling generates the non-universal terms via the asymmetric band, 
but as those parameters also play the role of the infrared cutoff in the
scaling the new fixed point cannot be reached.

In the last part, the tunneling impurity model with higher excitations
in the potential well is discussed. The question of physical
realization is not raised, it is only shown that the behavior in the
high energy region of the scaling can be modified by the electron-hole
symmetry breaking. This may change the parameters entering into the
low energy physics.  

\section{ELECTRON-HOLE SYMMETRY}
\label{sec2}

Treating the Kondo problem in the framework of the logarithmic
approximation, the following formulas are used for the energy integrals
\begin{equation}
\int\rho(\ep)\frac{n_F(\ep)}{\omega-\ep} d\ep\rightarrow 
\rho_0\log\frac{{\rm Max}(kT,|\omega|)}{D}\ ,
\end{equation}
and for $|\omega|\ll D$ 
\begin{equation}
\int\rho(\ep)
\frac{1}{\omega-\ep}d\ep\rightarrow 
\rho_0\int\limits_{-D}^D \frac{1}{\omega-\ep}d\ep\approx O 
\bigl (\frac{|\omega|}{D}\bigr )\rightarrow 0 \ ,
\end{equation}
where $\omega$ is the energy variable, $n_F$ is the Fermi distribution
function, and $\rho(\ep)$ is the conduction electron density of states
which is approximated by a constant $\rho_0$ inside the region
$-D<\ep <D$, $D$ being an energy cutoff, e.g. the bandwidth. Those
approximations do not affect the infrared divergent parts proportional
to $\log (\omega)$, but the non-diverging terms can strongly depend
on any deviation from those. Furthermore, the dependence on the wave
number $k(\ep)$ for an electron with energy $\ep$ is replaced by the
value at the Fermi energy $k_F$, thus
\begin{equation}
k(\ep)=k_F+\ep v_F\rightarrow k_F \ ,
\label{diszp}
\end{equation}
where $v_F$ is the Fermi velocity.
All of these approximations can be considered to be due to the 
electron-hole symmetry (e-h.s.).

That symmetry can be broken e.g. by introducing an energy dependent
density of states (see e.g. Ref.\onlinecite{Zarand})
\begin{equation}
\rho(\ep)=\rho_0 \bigl (1+\alpha \frac{\ep}{D}\bigr ) \ ,
\label{ehsb}
\end{equation}
where $\alpha$, in general, is of order of unity due to band structure
effects, or by using
different upper and lower cutoffs as $D_{up}$ and $D_{low}$ instead of
$D$. Another way\cite{Moustakas} to break the e-h.s. is to introduce a
local potential $V\delta ({\bf r})$ for the electron e.g. at the
impurity site ${\bf r}=0$.

The generally accepted approximation has been suggested by 
Haldane\cite{Haldene} for the case with different $D_{up}\neq D_{low}$ as
\begin{equation}
\log\frac{D_{up}}{|\omega|} +
\log\frac{D_{low}}{|\omega|} =
2\log\frac{(D_{low} D_{up})^{1/2}}{|\omega|}
\end{equation}
which only slightly affects the different quantities due to the
modified cutoff $D\rightarrow (D_{low} D_{up})^{1/2}$.

\section{KONDO PROBLEM WITH ASYMMETRIC CUTOFF ($D_{up}\neq D_{low}$)}
\label{sec3}

\begin{figure}
\centerline{\includegraphics[width=5in]{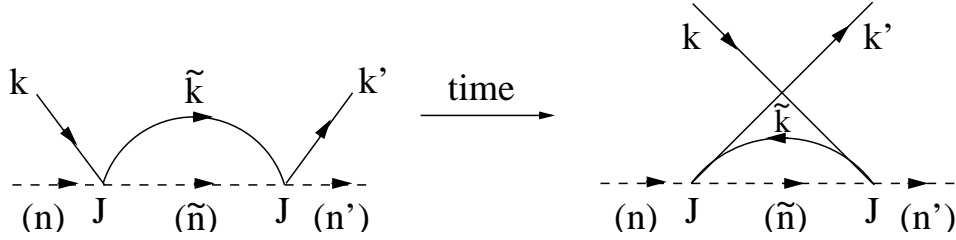}}
%
\caption{The basic two logarithmic time ordered diagrams are shown
  where the solid lines stand for the electrons and the dashed ones
  for the impurity spins. (In the parenthesis $n,n',{\tilde n}$ label
  the excited states in the tunneling problem.)}
\label{fig1}
\end{figure}
\begin{figure}
\centerline{\includegraphics[height=2in]{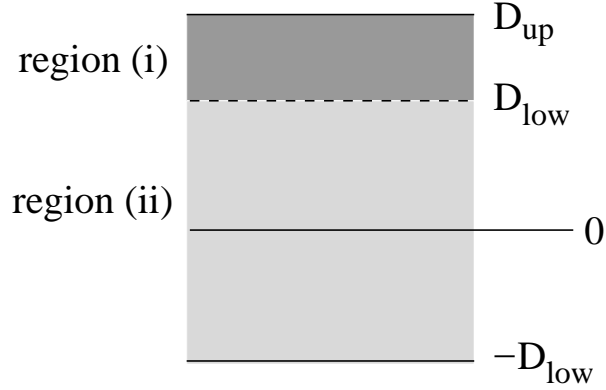}}
%
\caption{The energy regions (i) and (ii) due to the asymmetric
bandwidth $D_{up}\neq D_{low}$.}
\label{fig2}
\end{figure}

The magnetic Kondo problem is described by the Hamiltonian
$H=H_0+H_I$, where
\begin{equation}
H_0=\sum\limits_{{\bf k},\sigma} \ep({\bf k}) a_{{\bf
    k}\sigma}^\dagger a_{{\bf k}\sigma}
\end{equation}
and $a_{{\bf k}\sigma}^\dagger$ is the creation operator for an
electron with momentum ${\bf k}$, spin $\sigma$ and energy $\ep ({\bf
  k})$. Furthermore 
\begin{equation}
H_I=\sum\limits_{\scriptstyle i=x,y,z\atop\alpha\beta} 
J^i\Psi^\dagger_{\alpha} (0)
\sigma^i_{\alpha\beta}\Psi_\beta (0) S^i +
V\sum\limits_{\alpha}\Psi^\dagger_\alpha (0)\Psi_\alpha (0) \ ,
\label{HI}
\end{equation}
where $\Psi_\alpha (0)$ is the electron field operator at the
impurity site ${\bf r}=0$ with spin $\sigma$, $J^i$ is an anisotropic
exchange coupling, $S^i$ is the spin operator of the impurity,
and V is a static potential at the impurity site. For
the most of the cases $J^i=J$ and $V=0$ is taken as the initial values. 
The two basic logarithmic diagrams are shown in Fig.~\ref{fig1}. These
contribute as $\log\frac{D_{up}}{|\omega|}$ and
$\log\frac{D_{low}}{|\omega|}$ with different spin factors $\pm
[S(S+1)\mp \bsigma{\bf S}]$ where ${\bf S}$ is the spin vector
operator of the impurity and $\bsigma$ stands for the vector of Pauli spin
matrices of the conduction electrons. In case of $D_{up}=D_{low}$ the spin
diagonal terms corresponding to the diagrams of
Fig.~\ref{fig1} are canceling out.
That is not the case when we apply a poor man's
scaling in two steps for the case $D_{up}>D_{low}$\cite{VladarZawa} 
eliminating first
the energy region (i) $D_{up}>\ep>D_{low}$ and then (ii)
$D_{low}>\ep>-D_{low}$ (see Fig.~\ref{fig2}). 
In the region (i) only
the first diagram in
Fig.~\ref{fig1} must be considered, that is actually a one-electron
problem and a spin independent part is generated.  
The scaling equations for the couplings $J^i$ and $V$ have the 
form\cite{VladarZawa}
\begin{equation}
\frac{d}{dD}\left(
  \begin{array}[c]{c}
J^x\cr J^y\cr J^z\cr 2 V
  \end{array}
\right)=-\rho_0\frac{1}{2 D}\left(
  \begin{array}[c]{cccc}
2 V&-J^z &-J^y&J^x\cr -J^z&2 V&-J^x& J^y\cr -J^y&-J^x& 2 V& J^z\cr
J^x& J^y & J^z & 2 V
  \end{array}
\right)\left(
  \begin{array}[c]{c}
J^x\cr J^y\cr J^z\cr 2 V
  \end{array}
\right)
\end{equation}
The solutions of that equation for $J^x=J^y=J^z=J$ and for the initial
value $V=0$ are at the bottom of the region (i) ($D=D_{low}$)
\begin{equation}
J_{D=D_{low}}=\frac{J}{4}\biggl (\frac{3}{1-\frac{3}{2}
  J\rho_0\log\frac{D_{up}}{D_{low}}}+\frac{1}{1+\frac{1}{2}
  J\rho_0\log\frac{D_{up}}{D_{low}}}\biggr ) \ .
\end{equation}
There is also a potential $V$ generated, 
but the latter one does not have a role in
the symmetric region (ii). $J_{D=D_{low}}$ is the initial coupling for
the scaling in the region (ii) 
with cutoff $D_{low}$. In the extreme limit $D_{up}\gg
D_{low}$ the scaling in region (ii) does not lead to any essential
renormalization and the coupling diverges at the Kondo temperature
$T_K$, which is now 
\begin{equation}
T_K=D_{up} e^{-\frac{1}{\frac{3}{2} J\rho_0}}.
\end{equation}
Thus, having $D_{low}\ll D_{up}$ results in the change of the exponent
$(2 J\rho_0)^{-1}$ to $(3/2 J\rho_0)^{-1}$ and surprisingly $D_{low}$
drops out. That result is different from the expectation, that half of
the diagrams gives coupling with half strength, and, thus, the exponent 
should be $(J\rho_0)^{-1}$. 
In this result the generated potential scattering inside
region (i) plays a crucial role.

Such a limit may be interesting in the cases of degenerate
semiconductors with magnetic impurities or tunneling centers\cite{PbGeTe}.

\section{THE COMMUTATIVE TUNNELING MODEL E-H.S. BREAKING}
\label{sec4}

A two-level system (TLS) described by a $\bsigma_{TLS}$ Pauli
operator is coupled to the spherical electron waves with different
angular momenta $l$, but only to the azimuthal channel $m=0$ in case
where the axis of the TLS is chosen to be the $z$ axis\cite{Cox}.
Keeping only two angular momenta, the structure of the coupling can be
decomposed also into Pauli matrices. The simplest such Hamiltonian is
\begin{equation}
H_1=\sum\limits_{\scriptstyle i=x,y,z\atop{\alpha,\beta\atop k, k',
\sigma}} V^i\sigma^i_{TLS} 
a^\dagger_{k\alpha\sigma}(\sigma^i_{el})_{\alpha\beta} 
a_{k'\beta\sigma} +\Delta_0\sigma^x_{TLS}
\end{equation}
and the real spin $\sigma$ is a silent (conserved) variable. $V^z$
describes the screening coupling between the tunneling atom and the
electrons, $V^x$ and $V^y$ are due to electron assisted transitions
considering the TLS. The spontaneous tunneling rate is $\Delta_0$.
In the so-called commutative model $V^x=V^y=0$ and $V^z$ are not
renormalized, the renormalization affects only $\Delta_0$.

A. Moustakas and D. Fisher\cite{Moustakas} pointed out, that in such
a model the potential scattering at the impurity site (see the second
term on the right hand side of Eq.~(\ref{HI})) may induce $V^x\neq 0$, 
$V^y\neq 0$ due to $V^z\neq 0$ and $\Delta_0\neq 0$ at the initial
condition. That problem has been investigated in further details\cite{Zarand}.
In the original work\cite{Moustakas} bosonization is used which
clearly shows general tendencies, but it is hard to connect it to the
bare coupling constants, therefore, a systematic diagrammatic study
has been performed\cite{Zarand}.

\begin{figure}
\centerline{\includegraphics[width=4.5in]{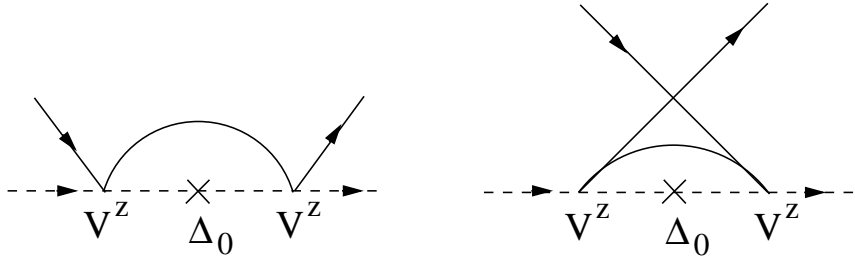}}
%
\caption{The diagrams generating the electron-TLS interaction which is
  off-diagonal in the TLS variables.}
\label{fig3}
\end{figure}
\begin{figure}
\centerline{\includegraphics[height=2.2in]{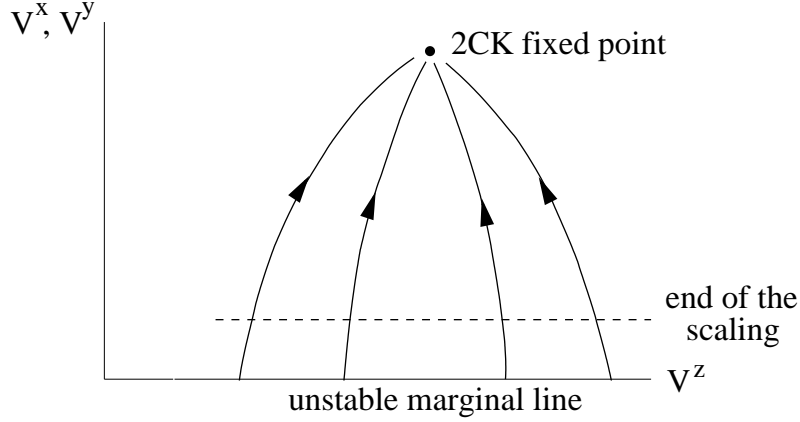}}
%
\caption{The scaling trajectories are shown. The marginal line
  $V^x=V^y=0$ is unstable due to the e-h.s. breaking and the
  trajectories are heading in the direction of the two-channel Kondo
  (2CK) fixed point. The low energy $\Delta_0$ restricts the
  scaling to the region below the dashed line.}
\label{fig4}
\end{figure}

The effect of the local potential scattering $V$ can be exactly taken
into account by solving the one-electron problem in the presence of
the potential, and then the e-h.s. breaking generates $\alpha\neq 0$ in
Eq.~(\ref{ehsb}). The $V^x$ coupling can be generated by the typical
diagrams shown in Fig.~\ref{fig3}. 
It is easy to show, that the two
diagrams in Fig.~\ref{fig3} cancel out if e-h.s exists, thus $\rho
(\ep)=\rho_0$. Using, however, $\rho (\ep)$ given by Eq.~(\ref{ehsb}),
$V^x$ is generated by the initial strength $\frac{\alpha}{D}
(V^z)^2\Delta_0$. One may expect that this non-commutative term can
drive the system away from the marginal line $V^z=const$,
$V^x=V^y=0$ to the strong coupling isotropic two-channel Kondo
fixed point, where $V^x=V^y=V^z\sim 1$. But that is not the case, as
$\Delta_0$ plays also the role of an infrared cutoff. Thus, $\Delta_0$
drives the couplings away from the marginal line and also stops it in
its original vicinity (see Fig.~\ref{fig4}). 
Thus, the $V^x=V^y=0$
model does not scale to
the interesting region with two-channel Kondo behavior with non-Fermi
liquid physics. In order to reach that, at least $V^x\neq 0$ with a
significant coupling strength is required.
Similar diagrams play a role in the next section.

\section{ORBITAL TWO-CHANNEL KONDO PROBLEM WITH CHANNEL SYMMETRY
BREAKING BY SPIN-ORBIT INTERACTION}
\label{sec5}

TLS's interacting with conduction electrons may represent the
two-channel Kondo problem\cite{Cox} where the channel degeneracy
is due to the real spin of the electrons. Motivated by the
discrepancies of the results of the two-channel Kondo problem and some
experimental data\cite{kiserletek}, here we examine the
possibility of breaking this degeneracy by the spin-orbit
interaction\cite{UZZ}.

\begin{figure}
\centerline{\includegraphics[height=2.2in]{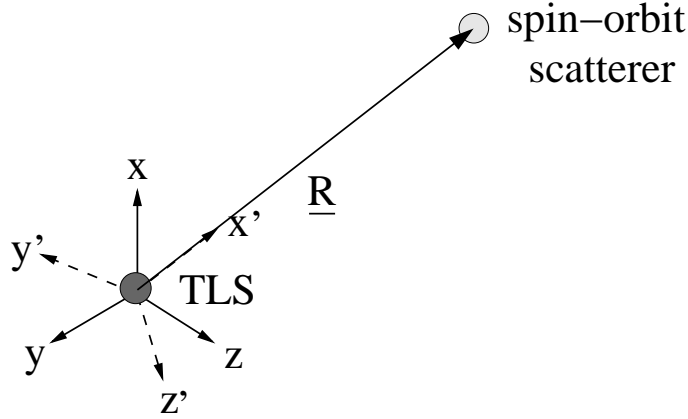}}
%
\caption{The TLS and the spin-orbit scatterer. The special $x,y,z$
  (for the TLS and electron orbital momentum spaces) and $x',y',z'$
  (for the electron spin space) coordinate systems with origin at the
  TLS are also indicated.}
\label{figo1}
\end{figure}

We consider a TLS interacting with conduction electrons which are also
interacting with a spin-orbit scatterer at a position ${\bf R}$
separated from the TLS (see Fig.~\ref{figo1}).
The corrections to the conduction electron Green's function due to the
interaction with the spin-orbit scatterer in appropriate coordinate
systems for the TLS/electron orbital momentum and for the electron
spin spaces (see Fig.~\ref{figo1}) can be summed up to infinite order, 
resulting in the following
change in the density of states of the conduction electrons\cite{UZZ,UZ}
\begin{equation}
\frac{\delta\rho_{\bf R}(\omega)}{\rho_0}=\Lambda
  \sigma^z_{l'l}\sigma^z_{\sigma'\sigma} \ ,
\label{deltarho}
\end{equation}
where the correct dispersion given by Eq.~(\ref{diszp}) was used, i.e. in
the density of states $\alpha\neq 0$ (see Eq.~(\ref{ehsb})).
$\Lambda$ depends on the strength of the spin-orbit interaction,
the energy and the corresponding hybridization matrix element of the
orbital where the spin-orbit interaction takes place, and the position
${\bf R}$ of the spin-orbit scatterer according to the TLS. The $l,l'$
and $\sigma, \sigma'$ indices correspond to the orbital momentum and
the real spin of the conduction electrons, respectively.
That treatment shows strong resemblance to the theory where the
spin-orbit scattering generates a surface anisotropy for a magnetic
impurity nearby the surface\cite{UZ}.

Redefining the dimensionless TLS-conduction electron couplings
($v^i=\rho_0 V^i$) by multiplying the bare couplings by the 
spin-dependent density of
states (see Eq.~(\ref{deltarho})), apart from the change
$v^x, v^y, v^z\rightarrow {\tilde v}^x, {\tilde v}^y,  {\tilde
  v}^z=v^z$, a new kind of coupling is generated\cite{UZZ} in a form 
$$\Lambda
v^z\sigma^z_{\sigma'\sigma}\sigma^z_{\alpha'\alpha}\delta_{l'l}$$
where the $\alpha,\alpha'$ indices label the TLS states.

Now we perform a scaling in the leading logarithmic approximation for
the general couplings $v^{\mu}_{\nu\rho}\sigma^{\mu}_{\alpha'\alpha}
\sigma^{\nu}_{l'l}\sigma^{\rho}_{\sigma'\sigma}$ where
$\mu,\nu,\rho=0,x,y,z$ according to the coordinate systems in 
Fig.~\ref{figo1}. 
To examine also the role of the e-h.s. breaking in the problem we used
Eq.~(\ref{ehsb}) for the conduction electron density of states in the
calculation. The leading logarithmic scaling equations are generated
by the diagrams in Fig.~\ref{fig1} and \ref{fig3} with the general
couplings at the vertices and the splitting $\Delta$ and the spontaneous
tunneling $\Delta_0$\cite{Cox} at the ``crosses'' on the TLS line.
As the scaling equations with the initial conditions are closed for
the subspace $\rho=0,z$, we can restrict the general equations to those
values and then define the ``average'' and the ``difference'' parts of
the couplings as
\begin{eqnarray}
v^\mu_\nu&:=&\frac{v^\mu_{\nu\uparrow}+v^\mu_{\nu\downarrow}}{2}=
v^\mu_{\nu 0} \hskip2cm (\mbox{average}),\cr
\delta v^\mu_\nu&:=&\frac{v^\mu_{\nu\uparrow}-v^\mu_{\nu\downarrow}}{2}=
v^\mu_{\nu z} \hskip2cm (\mbox{difference})
\end{eqnarray}
where $v^\mu_{\nu\uparrow}$ and $v^\mu_{\nu\downarrow}$ are the
couplings for up and down electron spins, respectively.
The initial values at $D=D_0$ are $v^s_p(0)=\delta_{sp} v_s$,
$v^s_0(0)=0$, $v^0_p(0)=0$, $\delta v^z_0=\Lambda v^z$, and the other 
differences are zero ($s,p=x,y,z$). 

After linearization in the differences, the scaling equations
for the averages decouple from the others, and it turns out that in leading 
order in $\frac{\alpha\Delta}{D}$, $\frac{\alpha\Delta_0}{D}$ the 
solution for the averages is the usual one\cite{Cox}, except that couplings 
$v^s_0\sim \frac{\alpha\Delta}{D}, \frac{\alpha\Delta_0}{D}$  
are generated\cite{UZZ}. Solving the scaling 
equations for the differences also in first order in $\frac{\alpha\Delta}{D}$, 
$\frac{\alpha\Delta_0}{D}$, most of the differences remain 
zero, except $\delta v^z_0, \delta v^x_z, \delta v^y_z, \delta v^z_x, 
\delta v^z_y$
where the $x, y, z$ indices are according to our special coordinate
system\cite{UZZ} (see Fig.~\ref{figo1}).
From the solution of the equations for those couplings we could
conclude\cite{UZZ}, that 
while $\delta v^z_0$ remains unrenormalized during scaling, the couplings 
$\delta v^x_z, \delta v^y_z, \delta v^z_x, \delta v^z_y$ are growing
like the usual solution for the averages multiplied by 
$\frac{\alpha\Delta}{D}$, $\frac{\alpha\Delta_0}{D}$, but only if 
$\alpha\neq 0$, thus the e-h.s. is broken.   

Thus, due to the interaction of the conduction electrons with a spin-orbit 
scatterer in a position ${\bf R}$ according to the TLS, new, relevant,
real spin dependent (thus channel degeneracy breaking) couplings between 
TLS and conduction electrons are generated in the case of e-h.s. breaking.

The situation is similar to the case presented in Sec.~\ref{sec4}. 
The e-h.s. breaking breaks the channel symmetry. The crossover between
the 2CK and 1CK behavior, however, cannot be reached as the $\delta v$
terms contain the factor $\frac{\alpha\Delta}{D}$ or 
$\frac{\alpha\Delta_0}{D}$ which is very small. Thus, again the
channel symmetry breaking is driven by $\Delta$ or $\Delta_0$, but the
same quantities stop the scaling long before the crossover is
reached. Thus, this cannot be relevant in the interpretation of the
experimental data\cite{kiserletek}.

\section{ELECTRON ASSISTED TRANSITION OF A HEAVY PARTICLE BETWEEN 
TWO STATES}  
\label{sec6}

Considering the interaction between a TLS and electrons the strength
of the electron assisted transition is not strong enough to obtain a
Kondo temperature $T_K$ which is required to fit experimental data. It
was suggested\cite{ZarandZawa} that the electron assisted transition 
to excited states with energies above the two lowest levels may give an
essential contribution to the electron assisted transition amplitudes
$V^x$ and $V^y$, but keeping only a few excited state levels gives
unphysical results\cite{Aleiner}. 
Here it is only shown in which sense the argument\cite{Aleiner} 
can be sensitive to
e-h.s\cite{ZarZaw}. The energy levels of the motion
of the localized atom are denoted by $E_n$, their wave functions are
$\Phi_n$ and the interaction between the electrons and the atoms is
described by the Hamiltonian
\begin{equation}
H'=U\int\Phi^\dagger_{n'}(x)\Phi_n (x)\Psi^\dagger (x)\Psi (x)
b^\dagger_{n'} b_n dx \ ,
\end{equation}
where the local interaction strength is $U$ and the atom in orbital
$n$ of the potential well is created by the operator $b^\dagger_n$. 
Now again the two diagrams in Fig.~\ref{fig1} should be considered.
Consider, e.g., the first diagram. In order to study the scaling
equations, the derivative $D\frac{\partial}{\partial D}$ must be taken,
and only the intermediate states with energies $E_{\tilde n} <D$ are
contributing. Then the contribution arising from
the region of the cutoff $D$ is (assuming that the wave function of the
atom depends on $z$),
\begin{eqnarray}
&&\sum\limits_{\tilde n}\int dz\int dz' e^{i (k_z-{\tilde k}_z) z}
\Phi^*_{\tilde n}(z)\Phi_n(z)\times \cr
&&\hspace*{2.38cm} e^{i ({\tilde k}_z-k'_z) z'}
\Phi^*_{n'}(z')\Phi_{\tilde n}(z')\rho(D) \Theta(D-E_{\tilde n}) \ ,
\end{eqnarray}
where $\omega$ is the incoming energy and $T=0$.
As it has been pointed out by Aleiner \textit{et al.}\cite{Aleiner}, 
if $D$ is large
enough, then ${\tilde n}$ covers a large enough number of states that the
approximate sum rule due to approximate completeness of the states,
\begin{equation}
\sum\limits_{\scriptstyle {\tilde n}\atop(E_{\tilde n} <D)} \Phi_{\tilde n}
(z)\Phi_{\tilde n} (z')\sim \delta(z-z') \ ,
\label{sumrule}
\end{equation}
is satisfied and then the contribution is simplified to 
\begin{equation}
\rho(D)\int
dz e^{i (k_z-k'_z) z} \Phi^*_{n'} (z)\Phi_n (z).
\end{equation}
The second diagram has exactly the same contribution but $\rho(D)$ is
replaced by $-\rho(-D)$. Thus if $\rho(D)=\rho(-D)$ holds, there is
an exact cancellation. However, that cancellation in the
renormalization group equation completely relies on the holding of
e-h.s. That is again an example, where results can be essentially
affected by breaking the e-h.s. 

It is clear, however, that the e-h.s breaking terms do not lead
to infrared divergences and can be relevant only in the vicinity of
the unrenormalized cutoff. Moving further from the cutoff region the
e-h.s. is gradually restored. Thus, the cancellation takes place. The
sum of the two diagrams contribute again  in the region where the sum
rule given by Eq.~(\ref{sumrule}) does not hold, having only a few
levels labeled by $\tilde n$. Thus the e-h.s. breaking has an effect
only through the possible enhancement of the couplings entering in the
treatment of that second region. That enhancement may be essential for
the value of the Kondo temperature. 

\section*{ACKNOWLEDGMENTS}

Among many of our colleagues we would like to thank especially 
I. Affleck and B.L. Altshuler for discussing Sec.~\ref{sec3} and
Sec.~\ref{sec6}, respectively.
This work was supported by the OTKA Postdoctoral Fellowship D32819
(O.\'U.) and by Hungarian grants OTKA F030041 (G.Z.), T029813, T030240, and 
T29236.


\begin{thebibliography}{9}

\bibitem{Zarand} A. Zawadowski, G. Zar\'and, P. Nozi\'eres,
  K. Vlad\'ar, G.T. Zim\'anyi, Phys. Rev. {\bf B56}, 12947 (1997).
\bibitem{Moustakas} A. Moustakas and D. Fisher, Phys. Rev. {\bf B51},
  6908 (1995) and ibid. {\bf B53}, 4300 (1996).
\bibitem{Haldene} F.D.M. Haldene, J. Phys. C {\bf 11}, 5015 (1978).
\bibitem{VladarZawa} K. Vlad\'ar and A. Zawadowski (unpublished).
\bibitem{PbGeTe} S. Katayama, S. Maekawa, H. Fukuyama,
  J. Phys. Soc. Jpn. {\bf 50}, 694 (1987) and N.B. Brandt,
  S.V. Demishev, V.V. Moshchalkov, and S.M. Chudinov,
  Fiz. Tekh. Poluprovodn. {\bf 15}, 1834
  (1981) (Sov. Phys. Semicond. {\bf 15}, 1067 (1982)).
\bibitem{Cox} see for review D. Cox and A. Zawadowski, Advances in
  Physics {\bf 47}, 599 (1998). 
\bibitem{kiserletek} S.K. Upadhyay, R.N. Louie, and R.A. Buhrman,
  Phys. Rev. B {\bf 56}, 12033 (1997), where in point contacts the I(V)
  characteristics shows a low energy cutoff. 
\bibitem{UZZ} O. \'Ujs\'aghy, G. Zar\'and and A. Zawadowski (unpublished).
\bibitem{UZ} O. \'Ujs\'aghy, A. Zawadowski, and B.L. Gyorffy,
  Phys. Rev. Lett. {\bf 76}, 2378 (1996); O. \'Ujs\'aghy and
  A. Zawadowski, Phys. Rev. B {\bf 57}, 11598 (1998).
\bibitem{ZarandZawa} G. Zar\'and and A. Zawadowski,
  Phys. Rev. Lett. {\bf 72}, 542 (1994), Phys. Rev. B {\bf 50}, 932 (1994).
\bibitem{Aleiner} I.L. Aleiner, B.L. Altshuler, Y.M. Galperin,
  T.A. Shutenko, Phys. Rev. Lett. {\bf 86}, 2629 (2001).
\bibitem{ZarZaw} A. Zawadowski and G. Zar\'and, cond-mat/0009283.

\end{thebibliography}
\end{document}